\documentclass[aps,prb,reprint,longbibliography,citeautoscript,superscriptaddress,showpacs]{revtex4-1}

\usepackage{graphicx,epsfig,float}
\usepackage{amssymb} 
\usepackage{amsmath}

\begin{document}
\title{Energetics and Structure of Domain Wall Networks in Minimally Twisted Bilayer Graphene under Strain}

\author{Irina V. Lebedeva}
\email{liv\_ira@hotmail.com}
\affiliation{CIC nanoGUNE BRTA, Avenida de Tolosa 76, 
San Sebasti\'an 20018, Spain}
\author{Andrey M. Popov} 
\email{popov-isan@mail.ru}
\affiliation{Institute for Spectroscopy of Russian Academy of Sciences, Fizicheskaya Street 5, Troitsk, Moscow 108840, Russia}

\begin{abstract}
The parameters of the triangular domain wall network in bilayer graphene with a simultaneously twisted and biaxially stretched bottom layer are studied using the two-chain Frenkel-Kontorova model. It is demonstrated that if the graphene layers are free to rotate, they prefer to stay co-aligned upon stretching the bottom layer and the regular triangular network of tensile domain walls is formed upon the commensurate-incommensurate phase transition. If the angle between the layers is fixed, the regular triangular network of shear domain walls is observed at zero elongation of the bottom layer. Upon stretching the bottom layer, however, the domain walls transform into the tensile ones and the size of the commensurate domains decreases. We also show that the parameters of the isosceles triangular domain wall network in twisted bilayer graphene under shear strain can be determined through purely geometrical considerations. Experimental analysis of the orientation of domain walls and period of the triangular network would, on the one hand, contribute to understanding of the interlayer interaction of graphene layers, and, on the other hand, serve for detection of relative strains and rotation between the layers. Vice versa external strains can be used to control the parameters of the triangular domain wall network and, therefore, electronic properties of twisted bilayer graphene. 
\end{abstract}
\maketitle
\section{Introduction}

Stacking dislocations in bilayer graphene arising as domain walls between commensurate domains with the AB and BA stackings were initially predicted for the case of uniaxial strain applied to one of the layers \cite{Popov2011}. Since then networks of domain walls separating commensurate domains have been obsevered using various experimental methods \cite{Alden2013, Lin2013, Butz2014, Yankowitz2014, Kisslinger2015, Ju2015, Jiang2016, Jiang2018, Huang2018, Yoo2019}. The effect of domain walls on electronic \cite{Wright2011, Ju2015, Huang2018, Yoo2019, Vaezi2013, Zhang2013, Hattendorf2013, San-Jose2013, San-Jose2014, Lalmi2014, Benameur2015, Koshino2013, Efimkin2018, Gargiulo2018, Ramires2018, Rickhaus2018}, magnetic \cite{Kisslinger2015, Wijk2015, Rickhaus2018} and optical \cite{Gong2013} properties of graphene has been studied.
Unusual plasmon reflection at domain walls opens up possibilities to manipulate two-dimensional plasmons \cite{Jiang2016}.
Topologically protected helical states in domain wall networks of minimally twisted bilayer graphene \cite{Ju2015, Vaezi2013, San-Jose2013, Huang2018, Rickhaus2018, Yoo2019} provide a new avenue for applications in valleytronics \cite{San-Jose2013} (see also for review \cite{Ren2016}).

In spite of the  interesting electronic properties of domain wall networks in minimally twisted bilayer graphene and their possible applications, the energetics and structure of such systems have been poorly studied. The majority of the previous theoretical works on the structure and energetics of domain walls were devoted to isolated domain walls which do not cross \cite{Popov2011, Lin2013, Butz2014, Lebedev2015, Lebedeva2016, Lebedev2017, Dai2016}.   
The approach developed in these papers allowed to predict the commensurate-incommensurate phase transition \cite{Pokrovsky1978} related to the formation of the first domain wall in two-dimensional bilayer systems with one layer stretched uniaxially and another layer free, such as bilayer graphene  \cite{Popov2011, Lebedeva2016}, bilayer boron nitride \cite{Lebedev2015, Lebedeva2016} and graphene-boron nitride heterostructure \cite{Lebedev2017}.

Although it is in principle possible to study the structure of triangular domain wall networks by atomistic \cite{Wijk2015, Gargiulo2018} and multiscale \cite{Zhang2018} simulations, the number of atoms in the supercell grows rapidly with decreasing the angle of relative rotation of the layers. Consideration of twisted bilayer with the minimal relative rotation angle of about $0.2^\circ$, such that the period of the network is an order of magnitude greater than the domain wall width, has been achieved so far using these methods \cite{Gargiulo2018, Zhang2018} and only the case of pure relative rotation of the layers without external strain has been addressed.

In most of the experiments \cite{Alden2013, Lin2013, Butz2014, Kisslinger2015, Jiang2016, Yoo2019}, the sizes of the commensurate domains are much greater than the domain walls width.
Moreover, different structures of the domain wall network are observed even within the same sample \cite{Alden2013, Lin2013, Kisslinger2015, Jiang2016}. Therefore, there is a need of a model capable of describing the networks with large domains under different external loads. It should be mentioned that local motion of domain walls by the electric field of the scanning tunneling microscope tip \cite{Jiang2018} and by the action of the atomic force microscope tip \cite{Yankowitz2014} has been demonstrated recently. Application of a strain to one of the layers can open an alternative way to manipulate the structure of domain wall networks.

It has been proposed lately \cite{Lebedeva2019} that the energy and structure of domain wall networks in bilayer graphene with the domain size much greater than the domain wall width can be described analytically within the two-chain Frenkel-Kontorova model \cite{Bichoutskaia2006}. However, only the case of co-aligned layers with a one biaxially stretched layer has been considered \cite{Lebedeva2019}.  Here we extend this approach to domain wall networks in bilayer graphene with a simultaneuosly twisted and biaxially stretched bottom layer.

In the following, we give the theory for domain walls in bilayer graphene: approximation of the potential energy surface of interlayer interaction energy, model for the local structure and energetics of isolated domain walls and its extension for regular triangular domain networks. Then we apply the extended model to analyze the characteristics of the networks formed under biaxial stretching of the bottom layer in the cases of free and twisted upper layers. In Section IV we briefly discuss the cases of compression, bending and shear load. Finally the conclusions are summarized.  

\section{Theory}
\subsection{Interlayer Interaction Energy}
\begin{figure}
	\centering
	\includegraphics[width=0.5\textwidth]{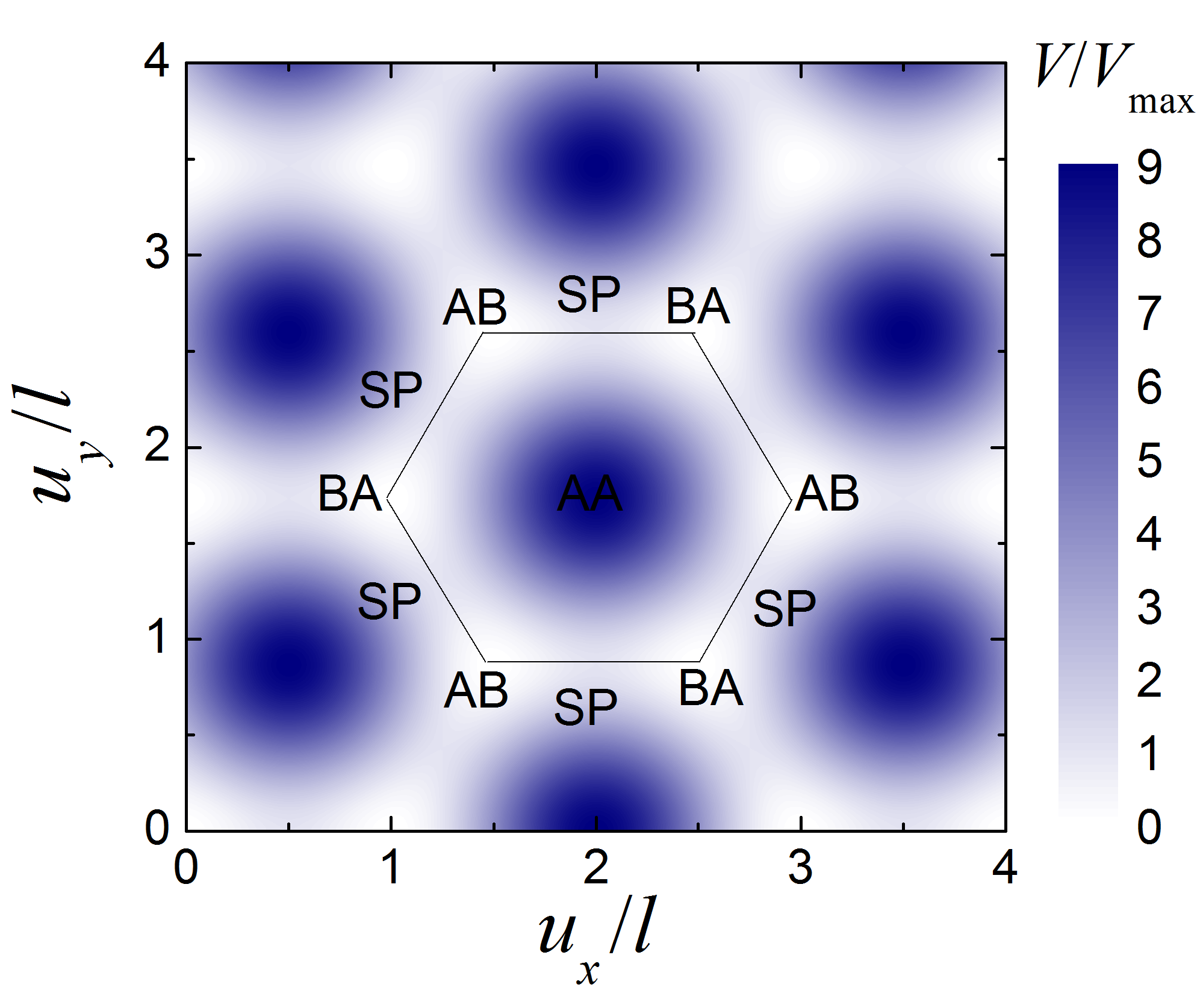}
\caption{Approximation of the interlayer interaction energy of bilayer graphene $V$ as a function of the relative displacements $u_x$ and $u_y$ of the layers along the armchair and zigzag directions, respectively, according to eq \ref{eq_n1}. The interlayer interaction energy $V$ is measured with respect to the AB (BA) stacking and divided by the barrier $V_\mathrm{max}$ to relative sliding of the layers, which corresponds to the relative energy of the SP stacking. The relative displacements $u_x$ and $u_y$ are given in units of the bond length $l$. The AB, BA, AA and SP stackings and the boundaries of the region of the potential energy surface spanned within a single dislocation node are indicated.} 
	\label{fig:pes}
\end{figure}

The structure and energetics of domain wall networks in bilayer graphene are determined by the potential energy surface of interlayer interaction energy for co-aligned layers. Density functional theory (DFT) calculations \cite{Popov2012,Reguzzoni2012,Lebedeva2011,Lebedeva2010,Lebedeva2011a} show that this potential energy surface, i.e. the dependence of the interlayer interaction energy on the relative in-plane displacement of the layers, can be approximated using the first Fourier harmonics (Figure \ref{fig:pes}):
\begin{equation} \label{eq_n1}
\begin{split}
V(u_x,u_y) = &V_0\Bigg(\frac{3}{2}+\cos\Big(2k_0u_x-\frac{2\pi}{3}\Big) \\
&-2\cos\Big(k_0u_x-\frac{\pi}{3}\Big)\cos\Big(k_0u_y\sqrt{3}\Big)\Bigg),
\end{split}
\end{equation} 
where $k_0$ is expressed through the bond length $l$ of graphene as $k_0 = 2\pi/(3l)$  and $u_x$ and $u_y$ are relative displacements of the layers in the armchair and zigzag directions, respectively. Bilayer graphene has two energetically degenerate but topologically inequivalent ground-state stackings, AB and BA, in both of which half of the atoms of the upper layer are on top of the centers of the hexagons and the other half on top of the atoms of the bottom layer. The relative displacement $\vec{u}=0$  in eq \ref{eq_n1} corresponds to the AB stacking. The structure and energy of domain walls depend only on the relative values of the interlayer interaction energy for different stackings and the energy in eq \ref{eq_n1}  is also given with respect to the AB (BA) stacking.

As seen from eq \ref{eq_n1} and Figure \ref{fig:pes}, the straight path between two adjacent minima AB and BA corresponds to the minimum energy path and the barrier to the displacement between the minima is reached in the middle of this path in the saddle-point (SP) stacking. The interlayer interaction energy grows fast upon deviation from the AB--SP--BA route compared to changes in the elastic energy of the layers. Therefore, the layers in domain walls are displaced along such paths \cite{Popov2011,Lebedeva2016,Alden2013}.

\begin{figure}
	\centering
	\includegraphics[width=0.5\textwidth]{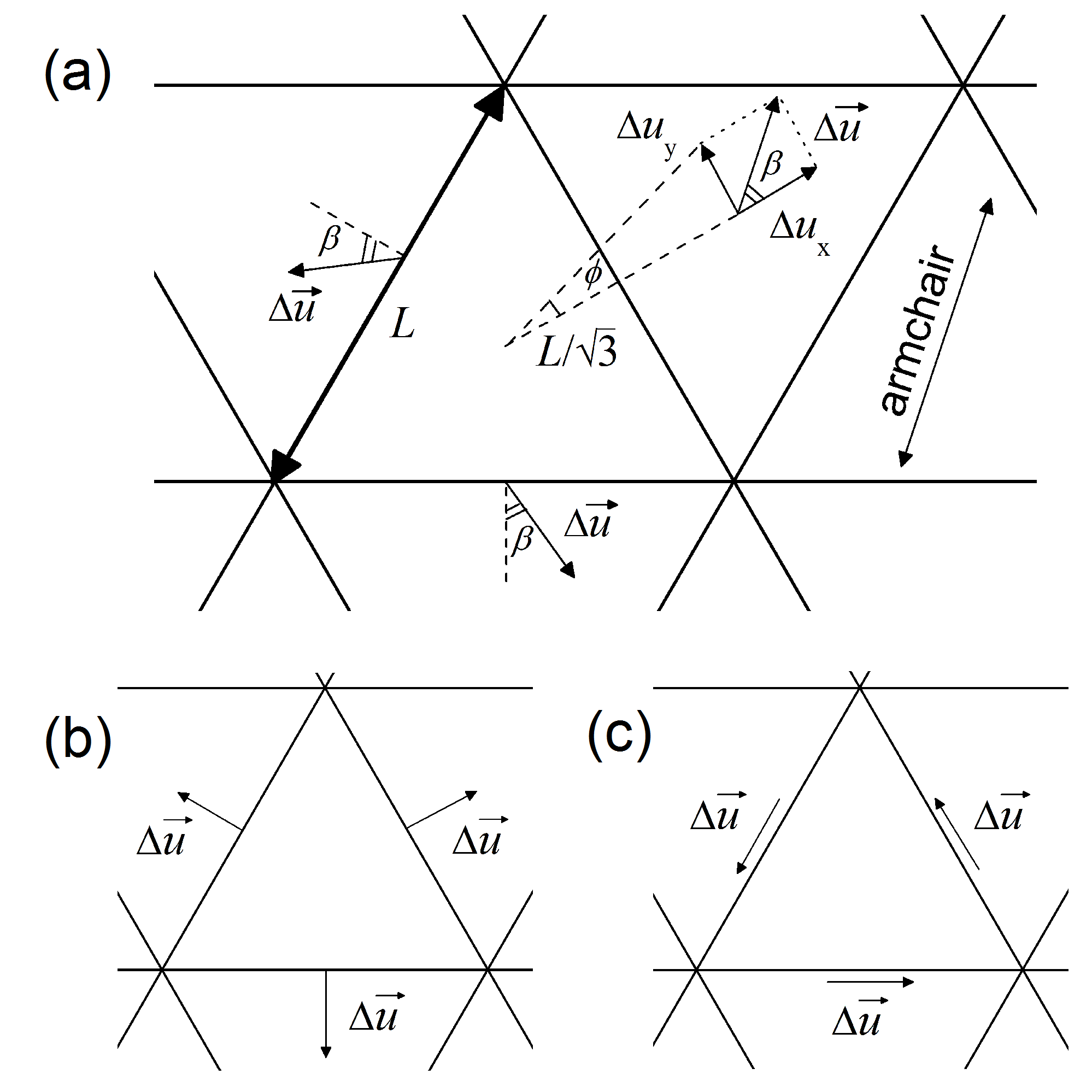}
	\caption{Schemes of regular triangular domain wall networks (black lines) in bilayer graphene: (a) general case, (b) tensile domain walls and (c) shear domain walls. The size $L$ of commensurate domains, changes $\Delta \vec{u}$  in the relative displacement of the layers in adjacent commensurate domains (equal to the Burgers vectors $\vec{b}$ of the domain walls up to a sign), components $\Delta u_x$ and $\Delta u_y$ of one of these vectors across (along the armchair direction) and along (along the zigzag direction) the domain wall, respectively, angles $\beta$ between the vectors $\Delta \vec{u}$ and normals to the domain walls and angle $\phi$ of the relative rotation of the layers are indicated.} 
	\label{fig:network}
\end{figure}

The Burgers vector $\vec{b}$  of a domain wall is related to the change $\Delta\vec{u}$ in the relative displacement of the layers in the commensurate domains separated by this wall as $\vec{b}=\pm \Delta\vec{u}$. The Burgers vectors of domain walls in graphene are thus aligned along the armchair directions and equal in magnitude to the bond length, $b=l$. The angle $\beta$ between the Burgers vector $\vec{b}$ and normal to the domain wall  (Figure \ref{fig:network}a) determines the character of the domain wall, which can change from tensile for the walls aligned in the zigzag direction ($\beta=0^\circ$, Figure \ref{fig:network}b) to shear for the walls aligned in the armchair direction ($\beta=90^\circ$, Figure \ref{fig:network}c).

It follows from eq \ref{eq_n1} that the dependence of the interlayer interaction energy on the displacement $u$ of graphene layers along the minimum energy path, which corresponds to the variation of the interlayer interaction energy across domain walls, can be written as
\begin{equation} \label{eq_n2}
\begin{split}
	V(u) = V_\mathrm{max}\left(2\cos{\left(k_0 u + \frac{2\pi}{3}\right)} +1\right)^2,
\end{split}
\end{equation}
where $V_\mathrm{max}=V_0/2$ is the barrier to relative sliding of the layers. This quantity is the key parameter of the Frenkel-Kontorova model describing the interlayer interaction. Unfortunately, the magnitude of the barrier $V_\mathrm{max}$ is not known with certainty. Because of the difficulty in description of long-range interactions in DFT, the corresponding values from literature lie in the wide range from 0.5 meV/atom to  2.1 meV/atom \cite{Kolmogorov2005,Reguzzoni2012,Aoki2007,Ershova2010,Lebedeva2011, Lebedeva2016a,Dion2004,Zhou2015} (in meV per atom of the upper/adsorbed layer, see Supporting Information). After comparison of a number of properties of bilayer graphene, graphite and boron nitride, such as shear and bulk moduli, shear mode frequencies, etc. for  various functionals corrected for van der Waals interactions (PBE-D2, PBE-D3, PBED3(BJ), PBE-TS, optPBE-vdW and vdW-DF2) with the experimental data, we previously came to the conclusion that the second version of the van der
Waals density functional (vdW-DF2) \cite{Lee2010} performs the best for the potential energy surface of these materials \cite{Lebedeva2016a}. Using the vdW-DF2 functional, we computed the barrier $V_\mathrm{max} = 1.61$~meV/atom. Other exchange-correlation functionals also gave the results in the range of 1.55--1.62 meV/atom when the interlayer spacing was fixed at the experimental one. The close estimate of 1.7 meV/atom was obtained from the experimental data on the shear mode frequencies in bilayer and few-layer graphene and graphite \cite{Popov2012}. However, a larger value of 
2.4 meV/atom was deduced from the experimental measurements of dislocation widths of various domain walls\cite{Alden2013}. Based on  these data, we assume that the error of our vdW-DF2 value for the barrier can reach 40\% and this is the main factor limiting the accuracy of our predictions for single domain walls and domain wall networks.

\subsection{Isolated Domain Walls}
To describe the energy and structure of domain walls in bilayer graphene analytically, we use the two-chain Frenkel-Kontorova model  \cite{Popov2011,  Lebedev2015, Lebedeva2016,Bichoutskaia2006, Popov2009, Lebedeva2019}. In this model, it is taken into account that both of the layers change their structure to accomodate domain walls. We, nevertheless, assume that the bilayer is supported \cite{Alden2013,Lin2013,Yankowitz2014} and neglect the out-of-plane buckling \cite{Butz2014,Lin2013}. Using the model, domain walls in double-walled carbon nanotubes \cite{Bichoutskaia2006, Popov2009},  bilayer graphene  \cite{Popov2011,Lebedeva2016, Lebedeva2017},  boron nitride \cite{Lebedev2015} and graphene-boron nitride heterostructure \cite{Lebedev2017} have been already investigated.

According to the two-chain Frenkel-Kontorova model, the energy related to formation of a single domain wall characterized by the angle $\beta$ between the Burgers vector and normal to the wall per unit length is given by \cite{Lebedev2015, Lebedeva2016}
\begin{equation} \label{eq_n3}
\begin{split}
\Delta W (\beta) = & \int\limits_{-\infty}^{+\infty} \bigg\{\frac{1}{4} K(\beta) \left|\frac{\mathrm d u}{\mathrm dx}\right|^2  + V(u)\bigg\}\mathrm{d}x 
\end{split}
\end{equation}
where coordinate $x$ corresponds to the direction perpendicular to the domain wall, $V(u)$ is the interlayer interaction energy per unit area of the bilayer along the minimum energy path between adjacent AB and BA minima given by eq \ref{eq_n2} and $K(\beta)=E \cos^2 \beta + G \sin^2 \beta$ describes the dependence of the elastic constant on the shear and tensile character of the domain wall \cite{Lebedev2015, Lebedeva2016}. Here $E=k/(1-\nu^2)$ and $G=k/2(1+\nu)$, where $\nu$ is the Poisson's ratio and $k$ is the elastic constant under uniaxial stress (determined by the Young's modulus $Y$ and thickness of graphene layers $h$ as $k=Yh$).

The condition $\delta \Delta W/ \delta u=0$ corresponds to the optimal relative displacement $u(x)$ that minimizes the formation energy of the domain wall in eq \ref{eq_n3}. Integration of this equation gives
\begin{equation} \label{eq_n4}
\begin{split}
\frac{1}{4}K(\beta) \left|\frac{\mathrm du}{\mathrm dx}\right|^2 = V(u). 
\end{split}
\end{equation}

The solution $u(x)$ of this equation is a soliton with a virtually constant slope near the center of the domain wall at $x=0$ \cite{Popov2011, Lebedev2015, Lebedeva2016,Bichoutskaia2006, Popov2009, Lebedeva2019}. Correspondingly, the characteristic width of domain walls referred to as a dislocation width  can be introduced as 
\begin{equation} \label{eq_n6}
	\begin{split}
		l_\mathrm{D}(\beta) = l \left|\frac{\mathrm{d} u}{\mathrm{d} x}\right|^{-1}_{x=0} =\frac{l}{2} \sqrt{\frac{K(\beta)}{V_\mathrm{max}}}.
	\end{split}
\end{equation}

From eqs \ref{eq_n2}, \ref{eq_n3} and \ref{eq_n4} it follows that the formation energy of domain walls per unit length is given by
\begin{equation} \label{eq_n7}
\begin{split}
\Delta W (\beta)= &\sqrt{K(\beta)} \int_0^l \sqrt{V(u)}\mathrm{d}u\\&=\sqrt{K(\beta) l^2 V_\mathrm{max}}\left(\frac{3\sqrt{3}}{\pi}-1\right).
\end{split}
\end{equation}

Using the parameters obtained by the DFT calculations \cite{Lebedeva2016} with the vdW-DF2 functional: $l = 1.430$~\AA, $k = 331 \pm 1$~J/m$^2$ and $\nu = 0.174 \pm 0.002$, we estimate that the dislocation width is 13.4 nm and 8.6 nm for tensile and shear domain walls, respectively. Because of the significant scatter in the data on the barrier $V_\mathrm{max}$ to relative sliding of the layers, as discussed in Section IIA, the accuracy of these estimates is about 20\%. The estimated dislocation widths are thus consistent with the experimental data \cite{Alden2013,Lin2013,Yankowitz2014} of 11 nm for tensile domain walls and 6 -- 7 nm for shear domain walls obtained for supported bilayer graphene. The deviation from the experimental data is comparable to that in the simulations based on the atomistic \cite{Gargiulo2018} and multiscale \cite{Zhang2018} approaches. The formation energy of domain walls per unit length computed using eq \ref{eq_n7} is 0.106 eV/\AA~and 0.068 eV/\AA~for tensile and shear domain walls, respectively.

\subsection{Domain Wall Networks}
\begin{figure}
	\centering
	\includegraphics[width=0.5\textwidth]{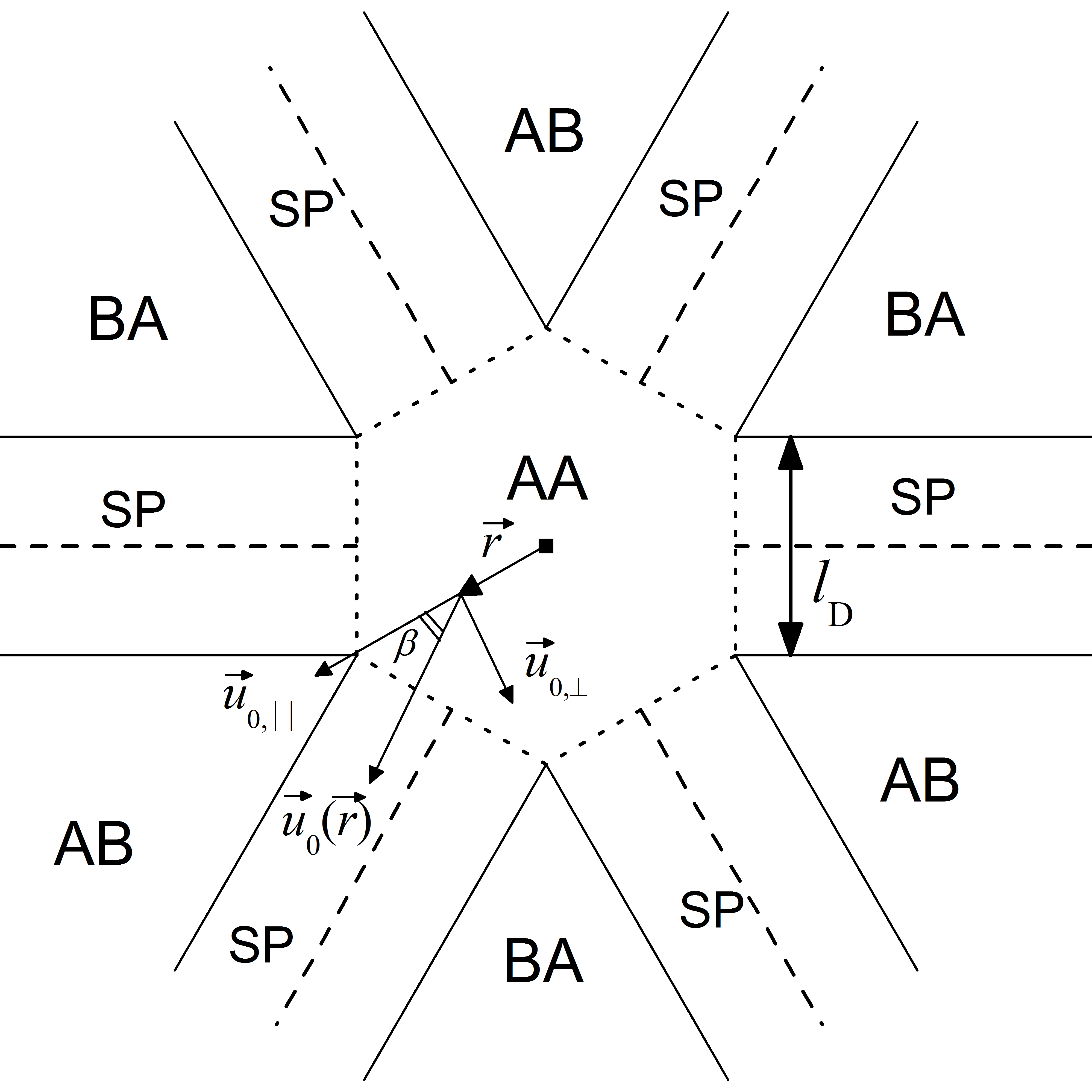}
	\caption{Scheme of a dislocation node in twisted bilayer graphene with a regular triangular domain wall network. Solid lines correspond to the boundaries between commensurate domains with the AB and BA stackings and domain walls. The center lines of the domain walls with the SP stacking are shown by dashed lines. The boundaries of the dislocation node with the AA stacking in the center are indicated by dotted lines. The dislocation width $l_\mathrm{D}$, relative displacement $\vec{u}_\mathrm{0}(\vec{r})$ of the layers at the position $\vec{r}$ with respect to the hexagon center ($|\vec{u}_\mathrm{0}(\vec{r})|=|\vec{r}| l/l_\mathrm{D}$), its components $\vec{u}_{0,\parallel}$ and $\vec{u}_{0,\perp}$ parallel and perpendicular to $\vec{r}$, respectively, and angle $\beta$ between the Burgers vectors and normals of the domain walls are shown. 
} 
	\label{fig:node}
\end{figure}

Let us now derive an expression for the energy of graphene bilayer with a domain wall network. If only an isotropic external load is applied, such as biaxial elongation of one of the layers and/or its relative rotation with respect to the other layer, it can be expected that all domain walls in the ground state of bilayer graphene, if any, should be characterized by the same angle $\beta$ between the Burgers vector and normal to the wall. Therefore, structures with a regular triangular domain wall network with six identical domain walls merging at each dislocation node should be considered as possible candidates for the ground state (Figures \ref{fig:network} and \ref{fig:node}).

Non-isotropic loads, such as uniaxial tensile or shear strain in one of the layers, should deform the equilateral triangular network and lead to formation of triangles with non-equal sides. Since the relative displacement of the layers accross two sides of a triangle occurs at an angle of 120$^\circ$ (Figures \ref{fig:pes} and \ref{fig:network}), the angle $\alpha_{12}$ between these sides is related to the angles $\beta_1$ and $\beta_2$ between the Burgers vectors and normals of the corresponding domain walls as 
\begin{equation} \label{eq_b}
\alpha_{12}=60^\circ+\beta_1-\beta_2.
\end{equation}
In the present paper, we mostly limit our consideration to the case of the isotropic load. Nevertheless, we also briefly discuss the structure of twisted bilayer under the special case of shear load providing triangular commensurate domains with two equal sides.

Let us consider a regular triangular domain wall network with the angle $\beta$ between the Burgers vectors and normals of the domain walls ($0\le\beta\le90^\circ$) and side $L$ of the triangles corresponding to the commensurate domains (Figure \ref{fig:network}). To use the Frenkel-Kontorova model, we assume that $L$ is much larger than the dislocation width, $L \gg l_\mathrm{D}$, where the dislocation width, $l_\mathrm{D}$, is determined by eq \ref{eq_n6}. 

Since the relative displacement of the layers grows by $l\sin{\beta}$ between centers of adjacent commensurate domains in the direction perpendicular to the line connecting them and the distance between them is $L/\sqrt{3}$ (Figure \ref{fig:network}), such a triangular network corresponds
 to the angle of relative rotation of the layers 
\begin{equation} \label{eq_n8_1}
\begin{split}
\phi\approx\tan{\phi} = \frac{l\sqrt{3} \sin{\beta}}{L}
\end{split}
\end{equation}
(for $\phi\lesssim10^\circ$). At the same time, the relative displacement of the layers  in the direction along the line connecting centers of adjacent commensurate domains  increases by $l\cos{\beta}$. Since the layers of the bilayer are made of the same material, this displacement is equally distributed between the layers and formation of the triangular network is associated with an extra relative biaxial elongation 
\begin{equation} \label{eq_n8_2}
\begin{split} 
\epsilon_0= \frac{\sqrt{3}l\cos{\beta}}{2L}
\end{split}
\end{equation}
in each of the layers. If the layers were free,  eqs \ref{eq_n3} and \ref{eq_n7} would describe the energy of domain walls in the bilayer graphene with the relative biaxial elongation of the bottom layer $\epsilon_0$ and relative rotation angle of the layers  $\phi$ with respect to the commensurate system with the zero biaxial elongation. As we consider the bottom layer with the relative biaxial elongation $\epsilon$,  eqs \ref{eq_n3} and \ref{eq_n7} correspond to the energy of domain walls with respect to the commensurate system with the relative biaxial elongation $\epsilon-\epsilon_0$. To compare the energies of the systems with the same elongation of the bottom layer, it is needed to substract the change of the elastic energy of the commensurate bilayer upon the change of the relative biaxial elongation from $\epsilon-\epsilon_0$ to $\epsilon$.

The energy of the bilayer with the domain wall network relative to the commensurate bilayer with the co-aligned layers and the same relative elongation $\epsilon$ of the bottom layer can thus be presented as
\begin{equation} \label{eq_n8}
\begin{split}
\Delta W_\mathrm{tot} = -\Delta W_\mathrm{el} + \Delta W_\mathrm{dw}+ \Delta W_\mathrm{dn},
\end{split}
\end{equation}
where $\Delta W_\mathrm{el}$ is the change of the elastic energy of the commensurate bilayer due to the extra elongation $\epsilon_0$, $\Delta W_\mathrm{dw}$ and $\Delta W_\mathrm{dn}$ are the contributions of the domain walls and dislocation nodes, respectively. We consider here the energies per unit area of the bilayer. 

The extra elongation $\epsilon_0$  corresponds to the increase in the elastic energy of the commensurate bilayer by 
\begin{equation} \label{eq_n9}
\begin{split}
 \Delta W_\mathrm{el} &=\frac{2k}{(1-\nu)}\left(\epsilon^2-\left(\epsilon-\epsilon_0\right)^2\right)\\&=\frac{2\sqrt{3}k\epsilon\cos{\beta}}{(1-\nu)}\frac{l}{L}-\frac{3k\cos^2{\beta}}{2(1-\nu)}\left(\frac{l}{L}\right)^2.
\end{split}
\end{equation}

The contribution of the domain walls is related to the energy $\Delta W$ per unit length of domain walls given by eq \ref{eq_n7} as $\Delta W_\mathrm{dw}=3L\Delta W/(2S)$, where $S=\sqrt{3}L^2/4$ is the area of one commensurate domain. Thus,
\begin{equation} \label{eq_n10}
\begin{split}
&\Delta W_\mathrm{dw}= \frac{2l}{L} \sqrt{3K(\beta) V_\mathrm{max}}\left(\frac{3\sqrt{3}}{\pi}-1\right)\\&=\frac{2l}{L} \sqrt{\frac{3kV_\mathrm{max}}{(1-\nu^2)}\left(\cos^2{\beta}+\frac{1-\nu}{2}\sin^2{\beta}\right)}\left(\frac{3\sqrt{3}}{\pi}-1\right)
\end{split}
\end{equation}

To estimate the contribution of the dislocation nodes, we suppose that the nodes have the shapes of hexagons with the side $l_\mathrm{D}$ equal to the width of domain walls and given by eq \ref{eq_n6} (Figure \ref{fig:node}). 
Since the relative displacement of the layers changes nearly linearly within domain walls \cite{Popov2011, Lebedev2015, Lebedeva2016,Bichoutskaia2006, Popov2009, Lebedeva2019}, we can consider the model in which the layers within a dislocation node are uniformly stretched and rotated, i.e. the relative displacement of the layers within the node is described as 
\begin{equation} \label{eq_n11}
\begin{split}
\vec{u}_\mathrm{0}(\vec{r})= l \left\{\sin{\beta}\frac{\vec{e}_z \times \vec{r}}{l_\mathrm{D}} + \cos{\beta} \frac{\vec{r}}{l_\mathrm{D}}\right\},
\end{split}
\end{equation}
where $\vec{e}_z$ is the unit normal to the graphene surface, $\vec{r}$ is the vector describing positions of the atoms within the dislocation node relative to the hexagon center and $\vec{u}_\mathrm{0}$ is chosen zero at the AA stacking. Correspondingly, the layers are in the AB and BA stackings at the vertices of the dislocation node and in the AA stacking at the center. As shown in our previous paper \cite{Lebedeva2019}, the assumptions we use for dislocation nodes correspond an error of 10--20\% in the formation energy. This is comparable to the error in estimates of the formation energy of domain walls related to the scatter of the available data on the barrier to relative sliding of graphene layers, $V_\mathrm{max}$ (see Section IIA).  This model is also consistent with the results of multiscale simulations \cite{Zhang2018}, where it was shown that the layers of twisted graphene are simply rotated with respect to each other within the dislocation nodes formed by shear domain walls ($\beta = 90^\circ$). The angle of rotation of the layers within the nodes determined in that paper was also close to $l/l_\mathrm{D}$. 

It follows from eq \ref{eq_n11} that the average elastic energy within a node is given by
\begin{equation} \label{eq_n11_1}
\begin{split}
V_\mathrm{el}=\frac{k\cos^2{\beta}}{2(1-\nu)}\left(\frac{l}{l_\mathrm{D}}\right)^2.
\end{split}
\end{equation}

The average energy of interlayer interaction within a node can be found as 
\begin{equation} \label{eq_n12}
\begin{split}
V_\mathrm{in} = \frac{\int_\mathrm{hex} V(u_x,u_y)\mathrm{d}u_x\mathrm{d}u_y}{\int_\mathrm{hex} \mathrm{d}u_x\mathrm{d}u_y}=\frac{3}{2}V_0=3V_\mathrm{max},
\end{split}
\end{equation}
where the energy is integrated over a hexagon of the potential energy surface 
with the center at the AA stacking and vertices at the AB and BA stackings  (see Figure \ref{fig:pes} and eq \ref{eq_n1}).  Note that $V_\mathrm{in}$ is the same as the average energy of interlayer interaction over the entire potential energy surface, i.e. the layers within a dislocation node in our model are fully incommensurate. Such a fully incommensurate system is observed when the layers are rotated by an angle at which the domain wall network disappears and no moir\'e pattern is formed \cite{Popov2012, Lebedeva2010, Lebedeva2011a}. And even at the angles corresponding to moir\'e patterns, the average interlayer interaction energy is almost the same as $V_\mathrm{in}$ (see Ref. \cite{Yakobson}).

The contribution of a single dislocation node to the relative energy of the system with a triangular domain wall network can then be presented as 
\begin{equation} \label{eq_n12_1}
\begin{split}
w_\mathrm{dn}=(V_\mathrm{el}+V_\mathrm{in})S_\mathrm{dn},
\end{split}
\end{equation}
where $S_\mathrm{dn}=3\sqrt{3}{l_\mathrm{D}}^2/2$ is the area of one dislocation node.  In the cases of tensile and shear domain walls, this quantity is 151 eV and 35 eV, respectively. For comparison, these values are equal to the formation energy of tensile and shear domain walls of length 144 nm and 52 nm, respectively.

Taking into account that the density of the nodes is $1/(2S)=2/(\sqrt{3}L^2)$, the contribution of the dislocation nodes to the energy of the bilayer with a triangular domain wall network relative to the commensurate state can be written as $\Delta W_\mathrm{dn} =2w_\mathrm{dn}/(\sqrt{3}L^2)$. Using eqs \ref{eq_n11_1} and \ref{eq_n12}, this gives
\begin{equation} \label{eq_n13}
\begin{split}
\Delta W_\mathrm{dn} = 3\left(\frac{l_\mathrm{D}}{L}\right)^2\left(3V_\mathrm{max} + \frac{k\cos^2{\beta}}{2(1-\nu)}\left(\frac{l}{l_\mathrm{D}}\right)^2\right).
\end{split}
\end{equation}

Based on eq \ref{eq_n6}, this contribution can be expressed as
\begin{equation} \label{eq_n14}
\begin{split}
\Delta W_\mathrm{dn} =3k \left(\frac{l}{L}\right)^2 \frac{2(5+2\nu)\cos^2{\beta}+3(1-\nu)\sin^2{\beta}}{8(1-\nu^2)}.
\end{split}
\end{equation}

Using eqs \ref{eq_n8}--\ref{eq_n10} and \ref{eq_n14}, the energy of the bilayer with a  triangular domain wall network relative to the commensurate system can be finally presented in the form
\begin{equation} \label{eq_n15}
\begin{split}
\Delta W_\mathrm{tot}= \frac{A(\beta)l}{L} +\frac{B(\beta)l^2}{L^2},
\end{split}
\end{equation}
where 
\begin{equation} \label{eq_n16}
\begin{split}
&A(\beta)=-\frac{2\sqrt{3} k\epsilon\cos{\beta}}{1-\nu}+ 2 \sqrt{\frac{3kV_\mathrm{max}}{(1-\nu^2)}}\\&\times\sqrt{\left(\cos^2{\beta}+\frac{1-\nu}{2}\sin^2{\beta}\right)}\left(\frac{3\sqrt{3}}{\pi}-1\right)
\end{split}
\end{equation}
and
\begin{equation} \label{eq_n17}
\begin{split}
B(\beta)=3k\frac{2(7+4\nu)\cos^2{\beta}+3(1-\nu)\sin^2{\beta}}{8(1-\nu^2)}.
\end{split}
\end{equation}

In the limit $L\to \infty$, $\Delta W_\mathrm{tot}$ in eq \ref{eq_n15} tends to zero as this case corresponds to the commensurate system. 

\section{Results}
\subsection{Free Upper Layer}
First let us consider the case when the upper layer can rotate freely with respect to the bottom layer. In this case, at $L=L_0$, where $L_0$ is the optimal period of the domain wall network, the following conditions should be fulfiled: $\partial \Delta W_\mathrm{tot}/\partial L=0$ and $\partial^2 \Delta W_\mathrm{tot}/\partial L^2\ge0$, where $\Delta W_\mathrm{tot}$ is given by eq \ref{eq_n15}. 

If $A(\beta)$ is positive, the optimal period of the network tends to infinity,  i.e. the commensurate state is energetically preferred over the systems with triangular domain wall networks characterized by a given $\beta$. If $A(\beta)$ is negative, the optimal state with a given $\beta$ corresponds to the bilayer with the domain wall network having the period
\begin{equation} \label{eq_n18}
\begin{split}
L_0(\beta) = -\frac{2B(\beta)}{A(\beta)}l
\end{split}
\end{equation}
and energy 
\begin{equation} \label{eq_n19}
\begin{split}
\Delta W_0(\beta) = -\frac{A^2(\beta)}{4B(\beta)}
\end{split}
\end{equation}
with respect to the commensurate system.

The critical relative biaxial elongation at which the network characterized by the angle $\beta$ becomes more energetically favourable than the commensurate state is determined by the condition $A(\beta)=0$, which gives
\begin{equation} \label{eq_n20}
\begin{split}
\epsilon_\mathrm{c}(\beta)&=\sqrt{\frac{V_\mathrm{max} (1-\nu)}{k(1+\nu)}\left(1+\frac{1-\nu}{2}\tan^2{\beta}\right)}\\&\times\left(\frac{3\sqrt{3}}{\pi}-1\right)=(1-\nu)\frac{\Delta W(\beta)}{kl\cos{\beta}}.
\end{split}
\end{equation}
As seen from this equation, the minimal critical elongation $\epsilon_\mathrm{c0}$ is reached for $\beta=0^\circ$: $\epsilon_\mathrm{c0}=\epsilon_\mathrm{c}(0)$. Therefore, at relative biaxial elongations $\epsilon > \epsilon_\mathrm{c0}=3.0\cdot 10^{-3}$, the ground state of graphene bilayer with a biaxially stretched bottom layer and free upper layer corresponds to the structure with a triangular domain wall network and  the commensurate-incommensurate phase transition takes place at $\epsilon = \epsilon_\mathrm{c0}$. Note that the expression for $\epsilon_\mathrm{c}(\beta)$ does not depend on $B$, i.e. the exact model used for dislocation nodes.  Because of the uncertainty in the value of the barrier $V_\mathrm{max}$ to relative sliding of the layers, as discussed in Section IIA, the accuracy of our estimate of the critical elongation is about 20\%.

As follows from eqs \ref{eq_n16} and \ref{eq_n19}, above the critical elongation $\epsilon_\mathrm{c}$, the relative energy of the bilayer with the domain wall network characterized by the angle $\beta$ can be written as 
\begin{equation} \label{eq_n21}
\begin{split}
\Delta W_0 (\beta)= -\frac{3k^2(\epsilon-\epsilon_\mathrm{c}(\beta))^2}{(1-\nu)^2} \frac{\cos^2{\beta}}{B(\beta)}.
\end{split}
\end{equation}
Eq \ref{eq_n17} shows that $\cos^2{\beta}/B(\beta)$ is maximal for $\beta=0^\circ$. Taking into account that for this $\beta$ the critical elongation $\epsilon_\mathrm{c}$ is also minimal, it is clear that 
formation of tensile domain walls aligned along the zigzag directions with $\beta=0^\circ$ is preferred over other types of domain walls for $\epsilon\ge \epsilon_\mathrm{c0}$. The regular triangular domain wall network with  $\beta=0^\circ$ corresponds to the zero relative rotation angle of the layers (see eq \ref{eq_n8_1} and Figure \ref{fig:network}b). Thus, if the upper layer of the bilayer is free, it stays co-aligned with the bottom layer upon the commensurate-incommensurate phase transition. 

It is seen from eqs \ref{eq_n16} -- \ref{eq_n18} for $\epsilon\ge \epsilon_\mathrm{c0}$ that the period $L_0(0)$ of the most energetically favourable domain wall network with $\beta=0^\circ$ is inversely proportional to the difference between the elongation of the bottom layer and critical elongation:
\begin{equation} \label{eq_n22}
\begin{split}
L_0(0)= \frac{\sqrt{3}(7+4\nu)}{4(1+\nu)(\epsilon-\epsilon_\mathrm{c0})}l\propto l(\epsilon-\epsilon_\mathrm{c0})^{-1}.
\end{split}
\end{equation}
The relative energy of the bilayer with such a domain wall network changes as
\begin{equation} \label{eq_n23}
\begin{split}
\Delta W_0 (0) = -\frac{4(1+\nu)k(\epsilon-\epsilon_\mathrm{c0} )^2}{(1-\nu)(7+4\nu)}\propto -kl^2/L_0^2(0),
\end{split}
\end{equation}
as follows from eq \ref{eq_n19}. Therefore, we can conclude that the commensurate-incommensurate phase transition in bilayer graphene taking place upon increasing the biaxaial elongation of the bottom layer is of the second order and can be described using the inverse period $L_0^{-1}(0)$ of the domain wall network as an order parameter. 

It should be pointed out that our model for bilayers with a triangular domain wall network is justified only for $L_0 \gg l_\mathrm{D}$. From eqs \ref{eq_n6} and  \ref{eq_n22}, it is clear that this condition is fulfiled for tensile domain walls with $\beta=0^\circ$ at elongations of the bottom layer
\begin{equation} \label{eq_n24}
\begin{split}
 \epsilon \ll \epsilon_\mathrm{max}=\epsilon_\mathrm{c0}+\frac{\sqrt{3}}{2}\left(7+4\nu\right)\sqrt{\frac{V_\mathrm{max} (1-\nu)}{k(1+\nu)}},
\end{split}
\end{equation}
i.e. $\epsilon \ll 3.3\cdot 10^{-2}$.

\subsection{Twisted Bilayer}
If the layers are rotated with respect to each other by the angle $\phi$ (Figure \ref{fig:network}a), eq \ref{eq_n8_1} describes the relation between the angle $\beta$ between the Burgers vectors and normals of the domain walls and the period $L$ of the triangular network. Then the first term in the relative energy of the system with a triangular domain wall network given by eq \ref{eq_n15} and determined by $A(\beta)$ from eq \ref{eq_n16} can be written using the notation $\chi=\cot{\beta}$ as
\begin{equation} \label{eq_n25}
\begin{split}
A(\beta)\frac{l}{L} = \left(A_0\sqrt{\chi^2+\frac{1-\nu}{2}}-A_1\chi\right),
\end{split}
\end{equation}
where
\begin{equation} \label{eq_n26}
\begin{split}
A_0=2\phi\sqrt{\frac{k V_\mathrm{max}}{(1-\nu^2)}}\left(\frac{3\sqrt{3}}{\pi}-1\right)
\end{split}
\end{equation}
and
\begin{equation} \label{eq_n27}
\begin{split}
A_1=2\phi \frac{k\epsilon}{(1-\nu)}.
\end{split}
\end{equation}

The second term in eq \ref{eq_n15}  determined by $B(\beta)$ from eq \ref{eq_n17} is given by
\begin{equation} \label{eq_n28}
\begin{split}
B(\beta)\left(\frac{l}{L}\right)^2=\left(B_0\chi^2+B_1\right),
\end{split}
\end{equation}
where
\begin{equation} \label{eq_n29}
\begin{split}
B_0=k\phi ^2\frac{7+4\nu}{4(1-\nu^2)}
\end{split}
\end{equation}
and
\begin{equation} \label{eq_n30}
\begin{split}
B_1=k\phi ^2\frac{3}{8(1+\nu)}.
\end{split}
\end{equation}
Note that $A_0$, $A_1$, $B_0$ and $B_1 \ge 0$. 

Finally eq \ref{eq_n15} takes the form
\begin{equation} \label{eq_n31}
\begin{split}
\Delta W_\mathrm{tot}= A_0\sqrt{\chi^2+\frac{1-\nu}{2}}-A_1\chi+B_0\chi^2+B_1.
\end{split}
\end{equation}

If there is no elongation applied to the bottom layer ($\epsilon=0$), then $A_1=0$. It is clear that in this case, the minimal energy of the triangular network is reached for $\chi=0$, i.e. shear domain walls aligned along the armchair directions with $\beta=90^\circ$ (Figure \ref{fig:network}c). The relation between the period $L_0=L$ of the triangular network and the angle $\phi$ of rotation of the layers in this case follows from the simple geometrical considerations  (see eq \ref{eq_n8_1}) and is given by
\begin{equation} \label{eq_n31_00}
\begin{split}
L_0=\frac{l\sqrt{3}}{\phi}. 
\end{split}
\end{equation}
For example, $L_0=142$ nm at $\phi=0.1^\circ$ and this value agrees well with the experimentally observed \cite{Yoo2019} period $L_\mathrm{exp}=120$ nm -- 140 nm at the same angle of rotation (note that the experimentally observed pattern is not really regular and the triangular commensurate domains are not exactly equilateral, probably due to the presence of inhomogeneous strains). For $\phi=0.4^\circ$, we get $L_0=35.5$ nm, while the experimental images \cite{Yoo2019} give $L_\mathrm{exp}=34$ nm -- 39 nm. 

\begin{figure}
	\centering
	\includegraphics[width=0.5\textwidth]{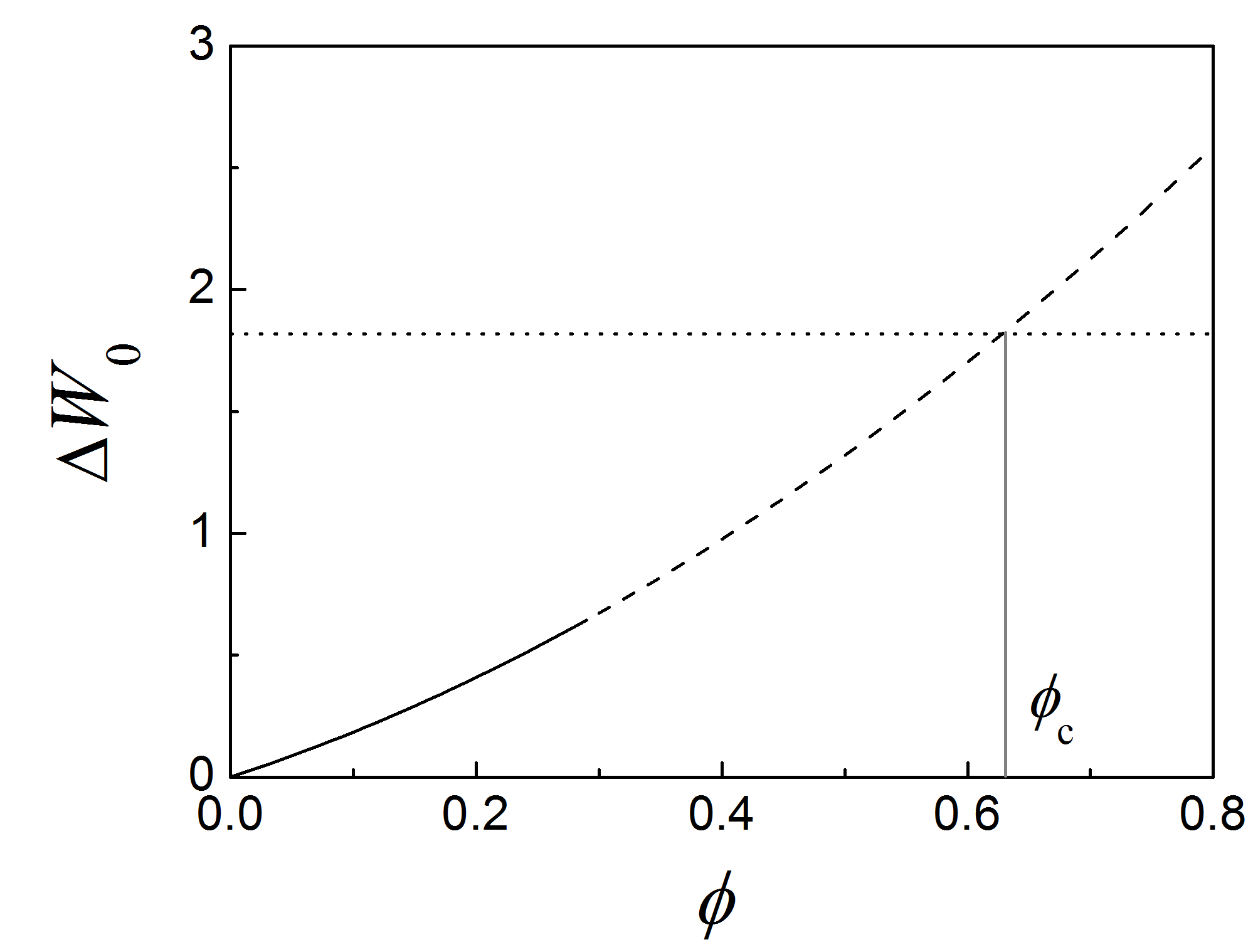}
	\caption{Estimated energy $\Delta W_0$ (in meV/\AA$^2$) of the twisted bilayer with the regular triangular domain wall network as a function of the angle $\phi$ (in degrees) of relative rotation of the layers with respect to the commensurate bilayer with co-aligned layers. The energy of the fully incommensurate state is shown by the horizontal dotted line. The estimated angle $\phi_{\mathrm{c}}$ at which the crossover to the  moir\'e structure takes place is indicated by the vertical line.} 
	\label{fig:rot}
\end{figure}

As follows from  eqs  \ref{eq_n26}, \ref{eq_n30} and \ref{eq_n31}, the energy of the twisted bilayer with the triangular domain wall network in the absence of the external strain applied grows upon increasing the angle $\phi$ of relative rotation of the layers with respect to the energy of the commensurate state of co-aligned layers as
\begin{equation} \label{eq_n31_0}
\begin{split}
\Delta W_0= \phi\sqrt{\frac{2 k V_\mathrm{max}}{(1+\nu)}}\left(\frac{3\sqrt{3}}{\pi}-1\right) + \frac{3k\phi ^2}{8(1+\nu)}.
\end{split}
\end{equation}
This function is plotted in Figure \ref{fig:rot}. The linear term in this expression is dominant for $\phi \ll 0.8^\circ$.

It can be estimated from eq \ref{eq_n31_0} that the relative energy $\Delta W_0$ of the twisted bilayer becomes comparable to the relative energy of the incommensurate moir\'e structure of about $V_\mathrm{in}$ (see eq \ref{eq_n12}) at $\phi=\phi_{\mathrm{c}}\sim0.6^\circ$.
A gradual crossover to the moir\'e structure was observed experimentally when the angle of relative rotation of the layers increased across the characteristic crossover angle $\phi_{\mathrm{c}}$ approximately equal to $1^\circ$ (Ref. \cite{Yoo2019}). A similar crossover angle was also obtained in atomistic \cite{Gargiulo2018} and multiscale \cite{Zhang2018} simulations. It is clear that in graphene layers rotated by about $1.1^\circ$, where superconductivity was discovered \cite{Cao2018}, the superstructure does not correspond to well-defined commensurate domains separated by soliton domain walls.

To define better the region of $\phi$ where our model is valid, we should consider the condition $L_0 \gg l_\mathrm{D}$ with the dislocation width $l_\mathrm{D}$ determined by eq \ref{eq_n6}. In the case of $\beta=90^\circ$, this means $\phi \ll \phi_\mathrm{max}=2 \sqrt{6V_\mathrm{max}(1+\nu)/k}=0.029 =1.6^\circ$. In atomistic simulations \cite{Gargiulo2018}, the dislocation width was independent of the period of the bilayer superstructure for $L_0 \gtrsim 5 l_\mathrm{D}$. This corresponds to $\phi \lesssim 0.33^\circ$.

Let us now consider the changes in the local structure of domain walls and period of the domain wall network upon application of non-zero elongation to the bottom layer ($\epsilon> 0$). In this case, the parameter $\chi$ of the optimal triangular domain wall network is determined by the condition $\partial \Delta W_\mathrm{tot}/\partial \chi =0$. As follows from eq \ref{eq_n31}, this condition is reduced to
\begin{equation} \label{eq_n32}
\begin{split}
\frac{\partial \Delta W_\mathrm{tot}}{\partial \chi}= \frac{A_0\chi}{\sqrt{\chi^2+\frac{1-\nu}{2}}}-A_1+2B_0\chi=0.
\end{split}
\end{equation}
Note that there is a unique solution for any $A_1 > 0$ and it corresponds to the minimum ($\partial^2 \Delta W_\mathrm{tot}/\partial \chi^2> 0$).

For $\chi \ll \sqrt{(1-\nu)/2}=0.64$ ($\tan{\beta} \gg \tan{57^\circ}$), the solution is
\begin{equation} \label{eq_n34}
\begin{split}
&\chi^{-1}\approx\chi_0^{-1}=\frac{2}{A_1}\left(B_0+\frac{A_0}{\sqrt{2\left({1-\nu}\right)}}\right) \\&=\frac{1}{\epsilon}\left(\epsilon_\mathrm{c0}\sqrt{\frac{2}{1-\nu}}+\phi\frac{7+4\nu}{4(1+\nu)}\right).
\end{split}
\end{equation}

In the opposite limit $\chi \gg \sqrt{(1-\nu)/2}$, eq \ref{eq_n32} gives
\begin{equation} \label{eq_n36}
\begin{split}
\chi^{-1}\approx\frac{2B_0}{A_1-A_0}=\frac{\phi(7+4\nu)}{4(1+\nu)(\epsilon-\epsilon_\mathrm{c0})}.
\end{split}
\end{equation}

In the intermediate region, the solution $\chi=\chi_{n \to\infty}$ can be found iteratively starting from eq \ref{eq_n34} through the expression
\begin{equation} \label{eq_n35}
\begin{split}
\chi_n^{-1}=\frac{2}{A_1}\left(B_0+\frac{A_0}{2\sqrt{\chi_{n-1}^2+\frac{1-\nu}{2}}}\right).
\end{split}
\end{equation}

\begin{figure}
	\centering
	\includegraphics[width=0.5\textwidth]{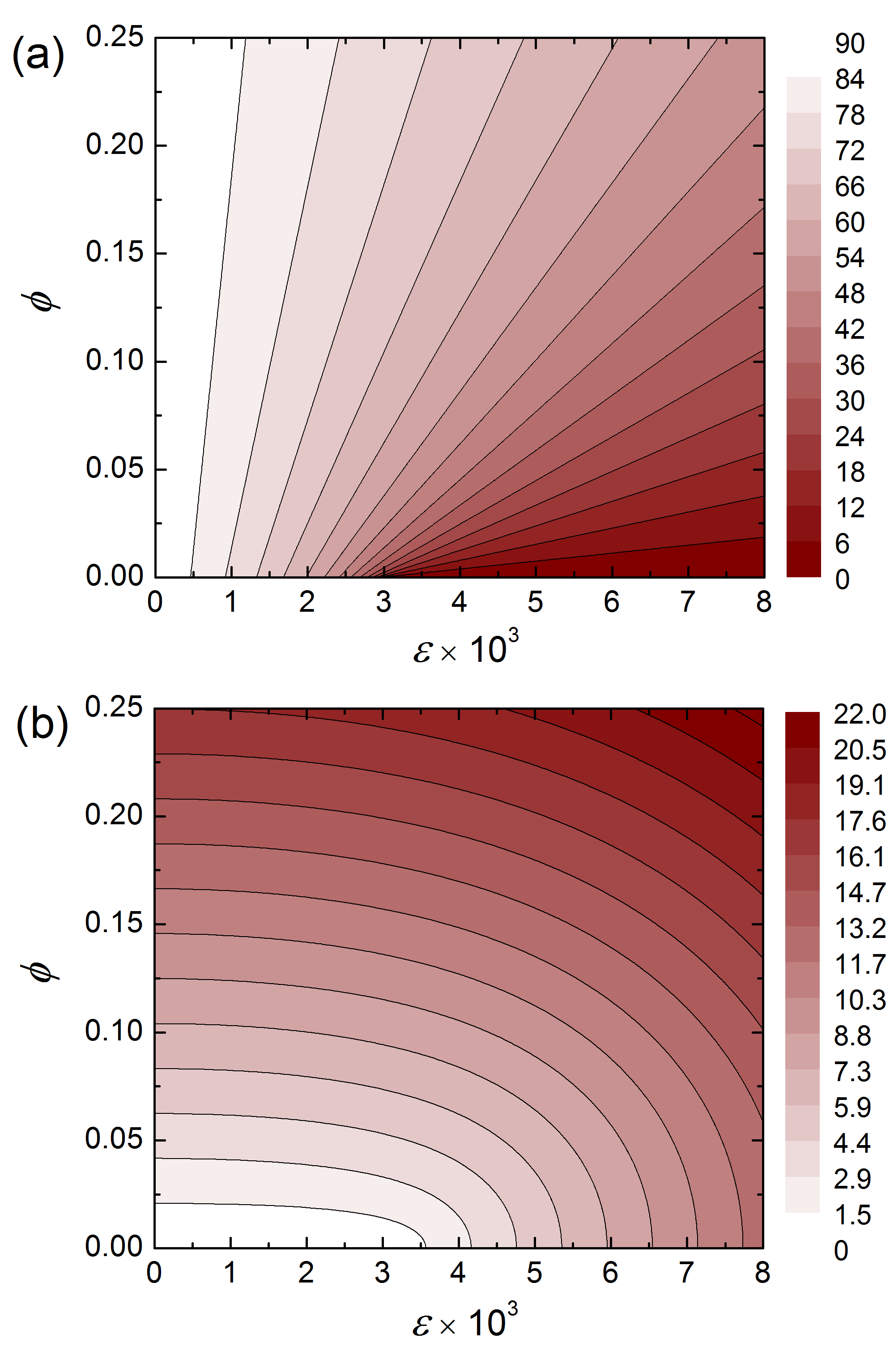}
	\caption{Estimated (a) angle $\beta$ (in degrees) between the Burgers vectors and normals of the domain walls and (b) inverse period $L_0^{-1}$ (in $\mu$m$^{-1}$) of the most energetically favourable triangular domain wall network in bilayer graphene as functions of the relative biaxial elongation $\epsilon$ of the bottom layer and angle $\phi$ (in degrees) of relative rotation of the layers. } 
	\label{fig:period}
\end{figure}

The angle $\beta=\arctan{\chi^{-1}}$ between the Burgers vectors and normals of the  domain walls and inverse period $L_0^{-1}=\phi/(l\sqrt{3}\sin\beta)$ of the most energetically favourable triangular network (see eq \ref{eq_n8_1}) computed using eq \ref{eq_n35} are shown in Figure \ref{fig:period}. It should be emphasized once again here  that our model is valid only for networks with the large period $L_0\gg l_\mathrm{D}$, where the dislocation width $l_\mathrm{D}\sim10$ nm is determined by eq \ref{eq_n6}, and we limit our consideration to such cases.  It is seen from Figure \ref{fig:period} that both the angle $\beta$ and period $L_0$ depend monotonically on the  angle $\phi$ of relative rotation of the layers and relative biaxial elongation $\epsilon$ of the bottom layer. Upon stretching the bottom layer, the angle $\beta$ decreases from 90$^\circ$ to 0$^\circ$, i.e. the domain walls rotate with respect to the commensurate domains so that their orientation changes from armchair to orthogonal zigzag and the character of the walls changes from shear to tensile. Increasing the angle $\phi$ of relative rotation of the layers has an opposite effect. The period $L_0$ of the network decreases, i.e. the commensurate domains shrink in size, upon increasing both $\phi$ and $\epsilon$.

Let us now discuss the limits of nearly shear and nearly tensile domain walls described by eqs \ref{eq_n34} and \ref{eq_n36}, respectively.  As seen from eq  \ref{eq_n34}, the condition $\chi \ll \sqrt{(1-\nu)/2}$ for the limit of nearly shear domain walls can be reformulated in terms of elongations as
\begin{equation} \label{eq_n35_3}
\begin{split}
\epsilon \ll \epsilon_\mathrm{l}=\epsilon_\mathrm{c0} + \phi\frac{(7+4\nu)\sqrt{1-\nu}}{4\sqrt{2}(1+\nu)}.
\end{split}
\end{equation}
Therefore, this limit also corresponds to the limit of small elongations. 

It also follows from eq \ref{eq_n34} that  in the limit of nearly shear domain walls,  $\tan \beta$  depends linearly on the angle $\phi$ of relative rotation of the layers and is inversely proportional to the relative biaxial elongation $\epsilon$ (Figure \ref{fig:period}a). The latter means that upon stretching the bottom layer at a given angle $\phi$, the domain walls should rotate away from the armchair direction at a virtually constant rate:
\begin{equation} \label{eq_n35_1}
\begin{split}
\frac{\mathrm{d}\beta}{\mathrm{d}\epsilon} \approx -\left(\epsilon_\mathrm{c0}\sqrt{\frac{2}{1-\nu}}+\phi\frac{7+4\nu}{4(1+\nu)}\right)^{-1}. 
\end{split}
\end{equation} 
The optimal period $L_0$ of the domain wall network in the limit of nearly shear domain walls  weakly depends on the elongation $\epsilon$ and changes inversely proportional to the angle $\phi$ according to eq \ref{eq_n31_00} (Figure \ref{fig:period}b).

The opposite limit of nearly tensile domain walls is described by  eq \ref{eq_n36}. As follows from the equation, this limit corresponds to the case $\epsilon > \epsilon_\mathrm{c0}$ and 
\begin{equation} \label{eq_n35_4}
\begin{split}
\phi \ll \phi_\mathrm{l} = \frac{4\sqrt{2}(1+\nu)}{(7+4\nu)\sqrt{1-\nu}}(\epsilon-\epsilon_\mathrm{c0}).
\end{split}
\end{equation}
Therefore, this is the limit of small angles $\phi$ of relative rotation of the layers at relative biaxial elongations $\epsilon$ exceeding the critical one. The function $\phi_\mathrm{l}(\epsilon)$ is inverse to $\epsilon_\mathrm{l}(\phi)$ (see eq \ref{eq_n35_3}). 

In the limit of nearly tensile domain walls, the angle $\beta$  between the Burgers vectors and normals of the  domain walls  is small and  approximately equal to $\chi^{-1}$. As seen from eq \ref{eq_n36}, this means that $\beta$ is proportional to  the angle $\phi$ and inversely proportional to the difference between the  elongation $\epsilon$ and its critical value $\epsilon_\mathrm{c0}$ (Figure \ref{fig:period}a). The rotation of the domain walls away from the zigzag direction thus occurs at a constant rate upon changing the angle $\phi$ at a given elongation $\epsilon$:
\begin{equation} \label{eq_n35_6}
\begin{split}
\frac{\mathrm{d}\beta}{\mathrm{d}\phi} \approx \frac{7+4\nu}{4(1+\nu)(\epsilon-\epsilon_\mathrm{c0})}.
\end{split}
\end{equation} 
The optimal period $L_0$ of the domain wall network  that follows from eq \ref{eq_n8_1} in this case is given by
\begin{equation} \label{eq_n37}
\begin{split}
L_0\approx\frac{2\sqrt{3}B_0l}{(A_1-A_0)\phi} =\frac{\sqrt{3}(7+4\nu)l}{4(1+\nu)(\epsilon-\epsilon_\mathrm{c0})}.
\end{split}
\end{equation}
Therefore, the period of the network in the limit of of nearly tensile domain walls does not depend on the angle $\phi$ of relative rotation of the layers and is determined only by the relative biaxial elongation $\epsilon$ of the bottom layer (Figure \ref{fig:period}b). Note that this equation is exactly the same as eq \ref{eq_n22} derived before for co-aligned graphene layers.

\section{Discussion}
Let us now discuss the superstructure of graphene bilayer under other types of mechanical load.

\subsection{Compression}
The same equations as for the bilayer with one stretched layer describe the system in which the layer is compressed. However, in the latter case, the strain can be effectively reduced nearly to zero through out-of-plane buckling. The crucial parameter for this process is the adhesion of the bilayer to the incommensurate substrate. When the elastic energy of the commensurate bilayer, $2k\epsilon^2/(1-\nu)$, becomes comparable to the binding energy $V_{sub}$, the bilayer can buckle away from the substrate losing in the adhesion to the substrate but gaining in the elastic energy. Thus, the compressive strain at which buckling out becomes energetically favourable for the commensurate bilayer can be estimated as $\epsilon_{sub}\sim\sqrt{(1-\nu)V_{sub}/(2k)}$. At such strains, there is no need in formation of domain walls. At $\epsilon\ll\epsilon_{sub}$, however, all the results obtained for the case of stretching should equally hold in the case of compression. 

For example, the magnitude of the binding energy for graphene on hexagonal boron nitride \cite{Sachs2011} or another graphene layer with the orientation corresponding to the fully incommensurate state \cite{Siahlo2017} is about $V_{sub} = 35$ meV/atom. This means that buckling out is supressed for such substrates till the compressive strain $\epsilon_{sub}\sim1.6\cdot 10^{-2}$. This is 5 times greater than the critical strain $\epsilon_\mathrm{c0}$ (see eq \ref{eq_n20}) and, therefore, it should be possible to observe the commensurate-incommensurate phase transition in co-aligned graphene layers or transformation of the domain wall network in twisted graphene bilayer not only upon stretching but also upon compression.

\subsection{Bending}
Bending of graphene bilayer can also give rise to formation of domain walls. In bended commensurate bilayer, one layer is slightly stretched and the other one slightly compressed. Therefore, it is necessary to include into the model that in the layers, there are tensile strains of opposite sign: $\pm\epsilon_R=\pm h_0/(2R)$, where $h_0=0.34$ nm is the interlayer spacing and $R \gg h_0$ is the curvature radius. Such strains couple to the relative displacement of the layers and reduce the formation energies of domain walls (eq \ref{eq_n7}) and dislocation nodes (eq \ref{eq_n12_1}). It can be expected that, similar to elongation applied to the bottom layer, bending of co-aligned layers can lead to formation of networks of tensile domain walls.  Bending of twisted bilayers should favour transformation of shear walls to tensile ones and reduce the period of the domain wall network. A modification of the Frenkel-Kontorova model is required to describe the effect of bending quantitatively and this will be performed elsewhere.

\subsection{Shear Strain}
\begin{figure}
	\centering
	\includegraphics[width=0.5\textwidth]{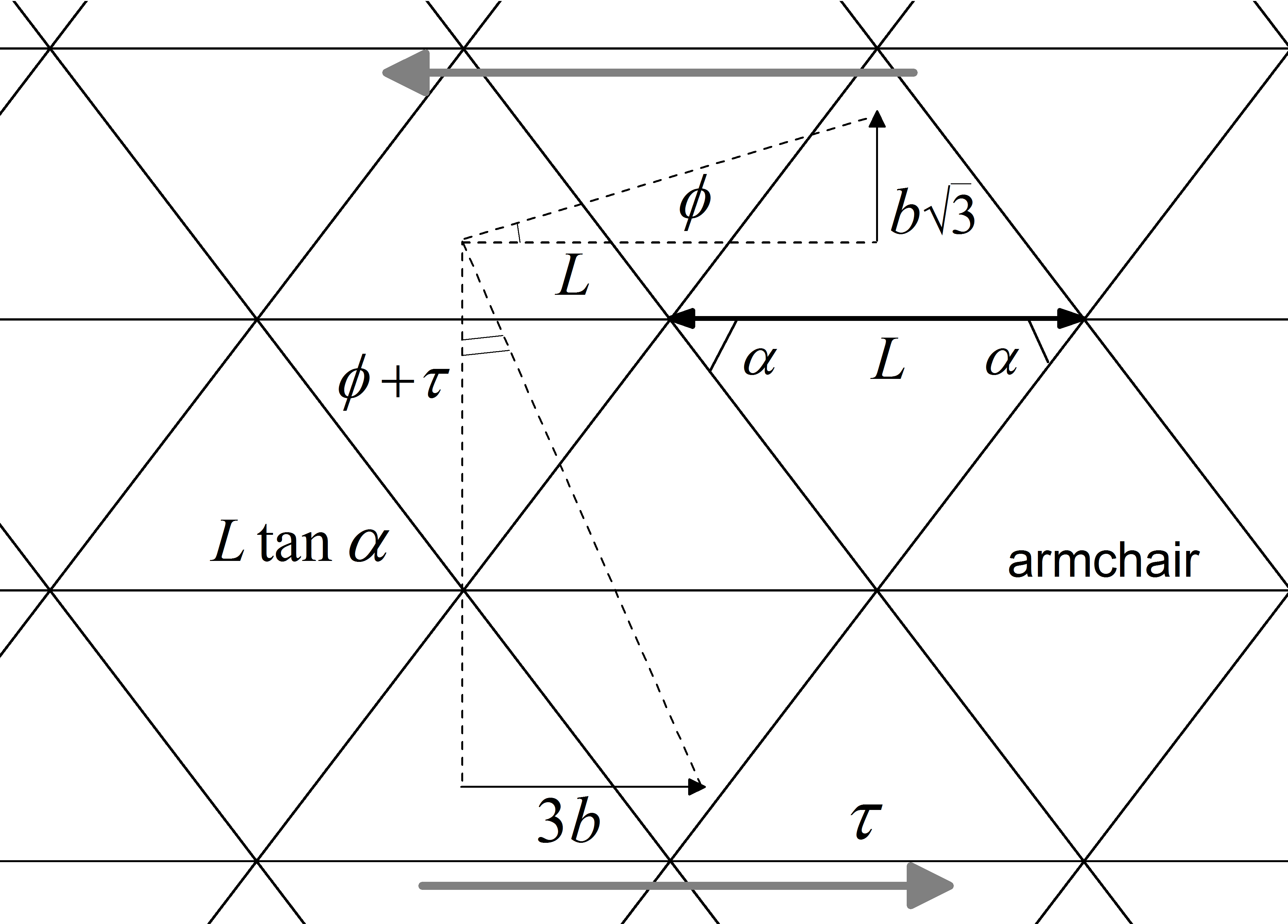}
	\caption{Scheme of the isosceles triangular domain wall network in twisted bilayer graphene with the ends of the bottom layer displaced in opposite armchair directions (as shown by the grey arrows). The length $L$ of the non-equal side of the triangles, changes in the relative displacement of the layers for the nearest equivalent commensurate domains in the orthogonal armchair and zigzag directions, angle $\phi$ of the relative rotation of the layers and shear strain $\tau$ applied to the bottom layer are indicated.} 
	\label{fig:shear}
\end{figure}

As discussed in Section IIIB, the period  of the triangular domain wall network in twisted graphene bilayer in the absence of elongation applied (Figure \ref{fig:network}c) is determined by the angle of relative rotation of the layers through purely geometrical considerations (see eqs \ref{eq_n8_1} and \ref{eq_n31_00}). The same also holds when additionally the ends of the bottom layer are displaced in opposite armchair directions so that shear strain $\tau$ is applied to the bottom layer, as shown in Figure \ref{fig:shear}. 
It can be expected that upon such an external load, the shear domain walls in the armchair direction parallel to the displacements of the ends of the bottom layer are preserved, while the other domain walls change their orientation in a symmetric way so that the triangular commensurate domains are left with two equal sides in the perpendicular zigzag direction.

Let us denote the length of the non-equal side along the armchair direction as $L$ and the equal angles of the triangles as $\alpha$. Since the non-equal side is formed by the shear domain wall, the angles between the Burgers vectors of the domain walls forming the two equal sides and their normals are $\beta = \alpha+30^\circ$ (see eq \ref{eq_b}).  We consider relative displacements of the layers in the equivalent commensurate domains with the same orientation. The distance between the nearest equivalent domains along the armchair direction is $L$ and the relative displacement of the layers changes between them by $b\sqrt{3}=l\sqrt{3}$ in the perpendicular zigzag direction. The shear strain applied does not contribute to this displacement and the rotation angle of the layers is thus given by $\phi\approx\tan{\phi}=  l\sqrt{3}/L$ (for $\phi\lesssim10^\circ$), the same as in the absence of the shear strain (see eq \ref{eq_n31_00}). The distance between the nearest domains in the perpendicular zigzag direction is $L\tan{\alpha}$ and both the shear strain and relative rotation of the layers contribute to the change of the relative displacement of the layers by $3b=3l$ so that $\tau+\phi\approx\tau+\tan{\phi}=3l/(L\tan{\alpha})$.
Therefore, $L=l\sqrt{3}/\phi$ and $\alpha=\arctan{\sqrt{3}\phi/(\phi+\tau)}$, i.e. the parameters of the isosceles triangular network are clear without energy optimization.

\section*{Conclusions}
Using the two-chain Frenkel-Kontorova model, we have studied the parameters of the triangular domain wall network for graphene bilayer with a simultaneously twisted and biaxially stretched bottom layer. We have focused on the case of the isotropic external load corresponding to equal elongations of the bottom layer along two orthogonal in-plane axes and relative rotation of the layers, when commensurate domains formed have the shape of equilateral triangles. 

We have demonstrated that if the layers are free to rotate, the layers stay co-aligned and formation of tensile domain walls is preferred upon stretching the bottom layer. In this case, the commensurate-incommensurate phase transition from the commensurate state to the incommensurate one with the regular triangular network of tensile domain walls aligned along the zigzag directions takes place at the critical relative biaxial elongation of the bottom layer of $3.0\cdot 10^{-3}$. 

If the angle between the layers is fixed, shear domain walls aligned along the armchair directions are observed at zero elongation of the bottom layer.  However, once the elongation is applied, the orientation of the domain walls changes from armchair to zigzag, i.e. the character of the walls changes from shear to tensile, and the period of the network decreases. The quantitative dependences of the optimal angle between the Burgers vectors and normals of the domain walls and period of the domain wall network have been obtained. It has been estimated that the formation energies of dislocation nodes are 151 eV and  35 eV in the cases of tensile and shear domain walls, respectively. 

When the ends of the bottom layer in twisted bilayer are shifted in opposite armchair directions, the triangular commensurate domains shrink or extend in the perpendicular zigzag direction and are left with only two equal sides. We have shown that the parameters of such a isosceles triangular domain wall network are determined by the relative rotation angle of the layers and shear strain applied through the purely geometrical considerations. 

Experimental  studies of the period of the triangular domain wall network by analogy with previous measurements using scanning tunneling microscopy \cite{Yankowitz2014}, scanning tunneling spectroscopy \cite{Huang2018}, transmission electron microscopy \cite{Alden2013, Lin2013,Yoo2019} or near-field infrared nanoscopy \cite{Jiang2016} can help to validate the \textit{ab initio}  results on the energy of the fully incommensurate state of graphene layers with respect to the commensurate state. At the same time, such measurements can provide a basis for detection of relative strains and rotation in graphene layers. 

External strains can be also applied to tune the parameters of the triangular domain wall network and, therefore, electronic \cite{Wright2011, Ju2015, Huang2018, Yoo2019, Vaezi2013, Zhang2013, Hattendorf2013, San-Jose2013, San-Jose2014, Lalmi2014, Benameur2015, Koshino2013, Efimkin2018, Gargiulo2018, Ramires2018, Rickhaus2018}, magnetic \cite{Kisslinger2015, Wijk2015, Rickhaus2018} and optical \cite{Gong2013} properties of twisted bilayer graphene. As we demonstrated, biaxial stretching of the bottom layer of twisted bilayer results in two structural effects: (1) a change of the character of the domain walls from shear to tensile and (2) a decrease of the period of the domain wall network. The calculations \cite{Koshino2013, San-Jose2014} show that at zero interlayer bias, tensile domain walls are almost insulating, while shear ones have only a soft transport gap. Therefore, stretching of the bottom layer of twisted bilayer should drive the system to a more pronounced insulating state with a larger transport gap. Under the interlayer bias applied, AB and BA regions of bilayer graphene correspond to two topological phases with opposite valley Chern numbers. As a result, domain walls separating AB and BA domains, which are insulating, confine one-dimensional conducting channels associated with topologically protected helical states \cite{Ju2015, Vaezi2013, San-Jose2013, Huang2018, Rickhaus2018, Yoo2019}. Because of the helicity, electrons can flow in these channels with no dissipation by momentum scattering. The topologically protected helical states arise in domain walls independent of their character, shear or tensile. Therefore, the effect of stretching on the electronic transport in twisted bilayer under the interlayer bias should be mostly related to the decrease in the period of the domain wall network. Upon stretching, the number of channels increases. However, their length decreases and there are more dislocation nodes, where mixing of currents from different channels occurs. The effect of these changes requires further investigation.

\section*{Acknowledgments}
AMP acknowledges the Russian Foundation for Basic Research (Grant 18-02-00985).

\section*{Supporting Information}
Tables of values of the barrier $V_\mathrm{max}$ to relative sliding of graphene layers available from first-principles calculations for bilayer graphene and graphite (PDF).

\bibliography{rsc}
\end{document}